\documentclass{PoS}

\usepackage{url}

\newcommand{\knn}{\mbox{\ensuremath{K\rightarrow \pi \nu \bar{\nu}}}}
\newcommand{\kpnn}{\mbox{\ensuremath{K^+\rightarrow \pi^+ \nu \bar{\nu}}}}
\newcommand{\knnn}{\mbox{\ensuremath{K_L\rightarrow \pi^0 \nu \bar{\nu}}}}
\newcommand{\kmtwo}{\mbox{\ensuremath{K^+\rightarrow \mu^+ \nu_{\mu}}}}
\newcommand{\kptwo}{\mbox{\ensuremath{K^+\rightarrow \pi^+ \pi^0}}}
\newcommand{\pme}{\mbox{\ensuremath{\pi \rightarrow \mu \rightarrow e}}}

\title{ORKA, The Golden Kaon Experiment: Precision measurement of $\kpnn$ and other rare processes}

\ShortTitle{ORKA, The Golden Kaon Experiment}

\author{\speaker{E.T. Worcester for the ORKA collaboration}\thanks{ASU, BNL, FNAL, 
Illinois, INFN-Napoli, INFN-Pisa,
INR-Moscow, JINR, Mississippi, Notre Dame,  
Texas-Arlington, Texas-Austin, TRIUMF, Tsinghua,
UBC, UNBC, UASLP}\\
        Brookhaven National Laboratory\\
        E-mail: \email{etw@bnl.gov}}

\abstract{ORKA is a proposed experiment to measure the $\kpnn$ branching ratio
with 5\% precision using the Fermilab Main Injector high-intensity
proton source. The detector design is based on the BNL E787/E949
experiments, which detected
seven $\kpnn$ candidate events. ORKA is expected to acheive two orders of 
magnitude improvement in sensitivity relative
to the BNL experiments as a result of enhancements to the beam line
and the detector acceptance. Precise measurement of the $\kpnn$ branching
ratio with the same level of uncertainty as the well-understood
Standard Model prediction
allows for sensitivity to new physics at and beyond the LHC mass scale.
Detector R\&D, simulation-based optimization of the experiment design,
and preparation of the experiment location are underway.}

\FullConference{2013 Kaon Physics International Conference,\\
		29 April-1 May 2013\\
		University of Michigan, Ann Arbor, Michigan - USA}

\begin{document}

\section{Introduction}
\label{sect:intro}
Precise measurements of rare kaon decay branching ratios
provide a window into potential new physics that is complementary
to that of direct searches at the energy frontier. Virtual effects
in $\kpnn$ loop diagrams allow us to
learn about the flavor- and CP-violating couplings of any new particles
that may be discovered at the LHC, and
to potentially observe the effects of
new particles with masses higher than those accessible by the LHC.
The $\kpnn$ branching ratio in the Standard Model is precisely
predicted by theory. The
technique to measure the branching ratio experimentally, though
challenging, is well-established, and, because seven candidate
$\kpnn$ events have 
already been seen, a large signal is assured. For these reasons,
precision measurement of the $\kpnn$ branching ratio is an ideal 
opportunity to test the Standard Model and search for new physics
beyond the Standard Model.

The ORKA collaboration proposes
to build a stopped-kaon experiment at FNAL to measure the $\kpnn$ branching
ratio with the same uncertainty as the Standard Model prediction,
using protons from the Main Injector.  
Section~\ref{sect:theory} describes the status of the Standard
Model calculation of the $\kpnn$ branching ratio, and the potential for
new physics to affect that branching ratio. The experimental history
and current status of the $\kpnn$ branching ratio measurement are described in
Sect.~\ref{sect:history}. Section~\ref{sect:expt} describes the BNL E787/E949
experiments on which the ORKA design is based, plans to dramatically
improve on the sensitivity of these experiments with ORKA, and the current 
status of the
ORKA experiment. Section~\ref{sect:summary} is a brief summary.

\section{$\kpnn$ in the Standard Model and Beyond}
\label{sect:theory}
\begin{figure}
\centering
\includegraphics[width=\columnwidth]{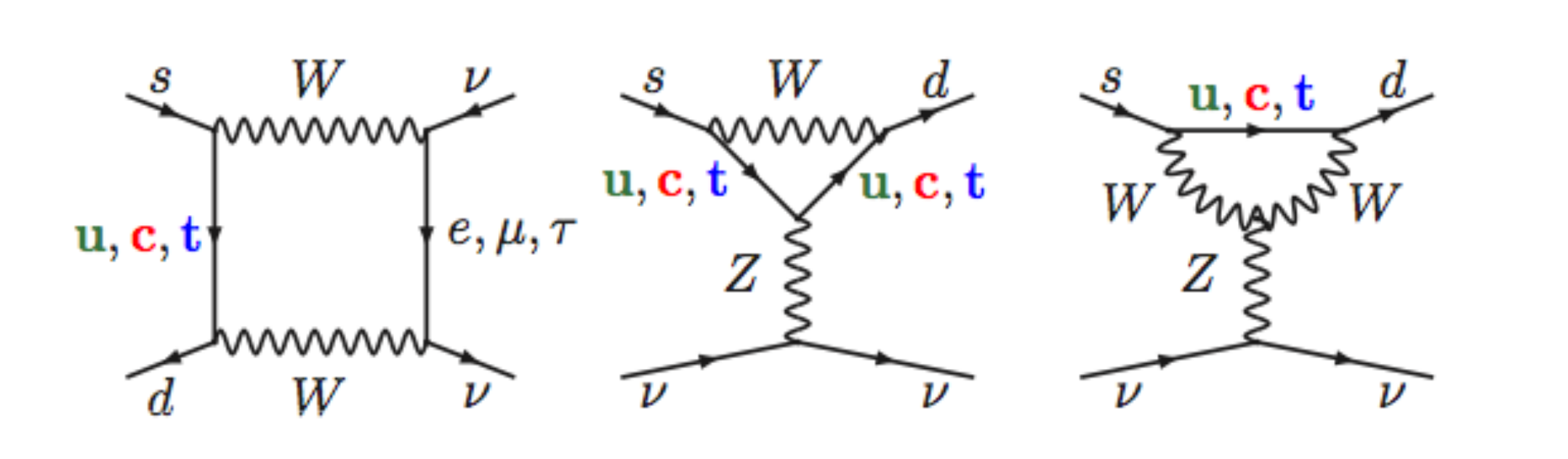}
\caption{One-loop diagrams for $\knn$ decay.}
\label{fig:loopdiagrams}
\end{figure}

The branching ratio of the rare $\kpnn$ decay is theoretically
well-understood. The one-loop diagrams for the $s \rightarrow d$ transition
are shown in Fig. \ref{fig:loopdiagrams}. The calculation is dominated
by the top-quark contribution while the charm-quark contribution is
significant but controlled.
The hadronic matrix element in the expression for
the $\kpnn$ decay rate is shared with that of the
$K^+ \rightarrow \pi^0 e^+ \nu_e$ decay, which allows for reliable
normalization. The current Standard Model (SM) prediction\cite{smpredict} is
\begin{equation}
\label{eq:smpredict}
B_{SM}(\kpnn) = (7.8 \pm 0.8) \times 10^{-11}.
\end{equation}
The $\sim$10\% uncertainty in Eq. \ref{eq:smpredict}
makes it one of the most precisely predicted 
flavor-changing-neutral-current decays involving quarks.
This uncertainty is dominated by
uncertainties in the CKM matrix; 
in the near future, improvements to the theory and measurements of the 
CKM elements
are expected to reduce it
by a factor of two. Therefore, deviation of
the measured decay rate from the SM prediction 
as small as 35\%  could be detected
with 5$\sigma$ significance for a measurement with 5\%
precision.

\begin{figure}
\centering
\includegraphics[width=0.9\columnwidth]{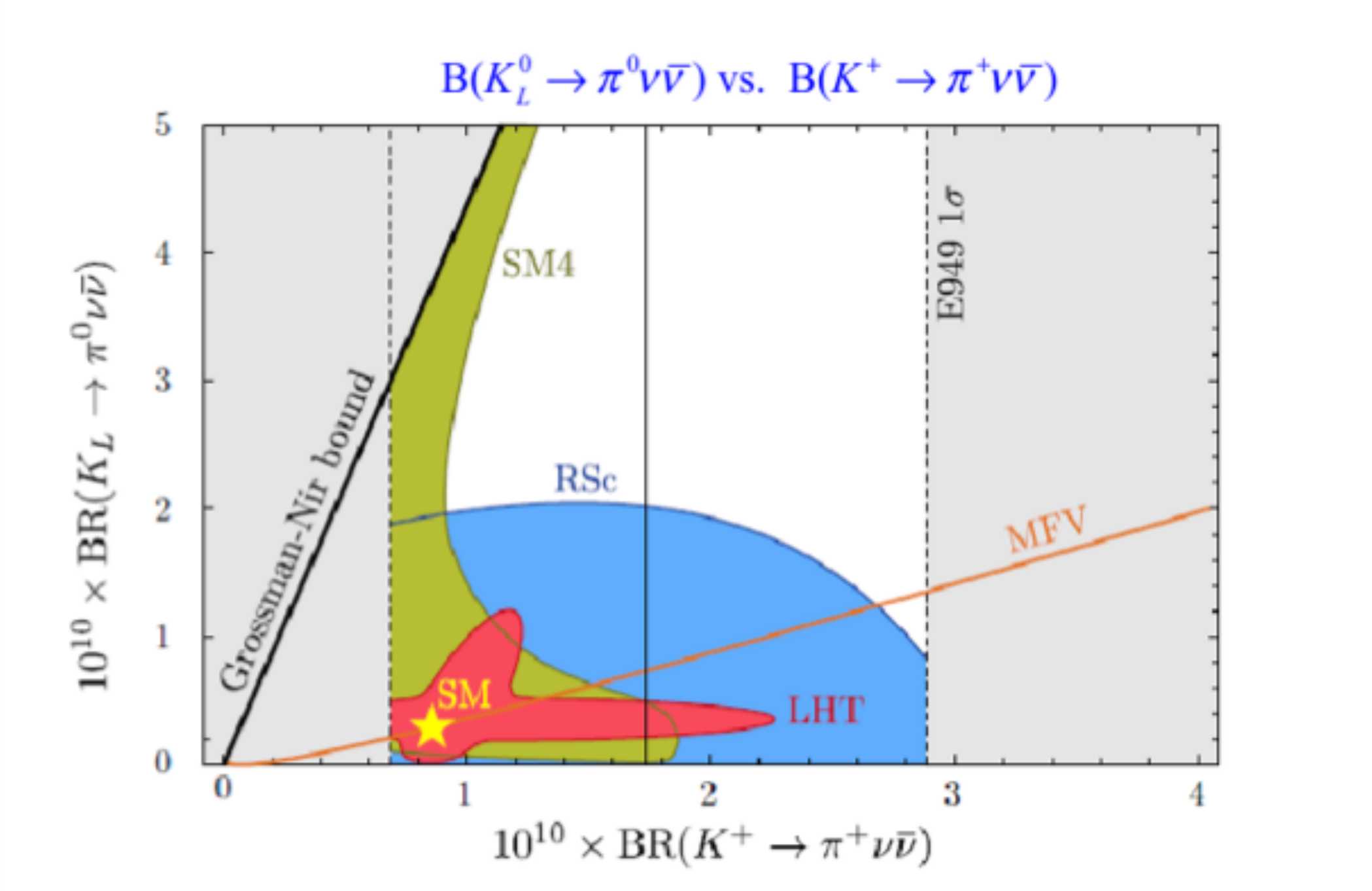}
\caption{
Figure courtesy of D.M. Straub\cite{Straub}.  Correlation
between the branching ratios of $\knnn$ and $\kpnn$ in minimal flavor
violation (MFV), and three concrete new physics models: 
Littlest Higgs with T parity (LHT)\cite{LHT}, Randall-Sundrum
with custodial protection (RSc)\cite{RSc}, and four quark generations 
(SM4)\cite{SM4}. The shaded area is ruled out experimentally at the 1$\sigma$
level or model-independently by the Grossman-Nir bound\cite{GNbound}.
The Standard Model point is marked by a star, and the central value of
the BNL E787/E949 result\cite{e949prd} is indicated by the solid, 
vertical line.}
\label{fig:newphysics}
\end{figure}

Many models of new physics predict
significant enhancements to the Standard Model branching ratios
of $\kpnn$ and $\knnn$. This is illustrated in Fig. \ref{fig:newphysics}, 
which shows the range of branching ratio predictions for minimal flavor
violation (MFV), Littlest Higgs with T parity (LHT)\cite{LHT}, Randall-Sundrum
with custodial protection (RSc)\cite{RSc}, and four quark generations 
(SM4)\cite{SM4}. It is interesting to note that tighter constraints on non-SM
effects for $\knnn$ may be imposed by 
direct CP violation ($\epsilon^{\prime}/\epsilon$), so that
less than a factor of two enhancement in this branching ratio would be
allowed\cite{Haisch_PXPS}. 

\section{$\kpnn$ Experimental History and Status}
\label{sect:history}
The first upper limit on the $\kpnn$ branching ratio was set by
a heavy-liquid bubble chamber experiment in 1969\cite{camerini}. The
first $\kpnn$ event was seen at BNL by E787\cite{e787first} in 1997.
BNL E949 represented a significant upgrade from BNL E787;
E949 detected four additional candidate events.
The current branching ratio measurement,
\begin{equation}
B(\kpnn) = (17.3^{+11.5}_{-10.5}) \times 10^{-11},
\end{equation}
is a combined result based on the seven candidate events observed by the 
BNL E787 and E949
experiments\cite{e949prd}. The central value of the measured branching
ratio is approximately twice that of the Standard Model
prediction, but the measurement and prediction are consistent.

All $\kpnn$ experiments to date have used stopped kaons. 
The NA-62\cite{na62_status13} 
experiment at CERN, currently under construction, will use a 
complementary
decay-in-flight technique. 
In contrast to the stopped-kaon experiments, which have greater
sensitivity at high $\pi^+$ momenta, the decay-in-flight technique is
more sensitive to $\kpnn$ decays with lower $\pi^+$ momenta.
NA-62 expects to see $\sim$100 total
$\kpnn$ events at the SM level and make a measurement
of the $\kpnn$ branching ratio with precision of $\sim$10\%. 

ORKA is a proposed experiment to
use the stopped-kaon technique and the FNAL Main Injector to detect
$\sim$1000 $\kpnn$ decays and make a measurement of
the $\kpnn$ branching ratio with $\sim$5\%
precision. The following section describes
the stopped-kaon technique and the proposed ORKA experiment. 

\section{The ORKA Experiment}
\label{sect:expt}
ORKA will use protons from the Main Injector (MI) at FNAL and will
build on the BNL E949 detector design to measure the $\kpnn$ branching
ratio with $\sim$5\% precision. Section \ref{sect:bnl} provides some
details on the stopped-kaon technique and analysis of the BNL E787/E949
data. Plans to significantly improve on the sensitivity of the BNL experiments
at ORKA are described in Sect. \ref{sect:orkaexpt}.
Section \ref{sect:otherphys} describes the breadth of physics measurements
that may be made using the ORKA experiment. The status of the ORKA experiment
is described in Sect. \ref{sect:orkastatus}.

\subsection{Stopped-Kaon Technique in BNL E787/E949}
\label{sect:bnl}
\begin{figure}
\centering
\includegraphics[width=0.47\columnwidth]{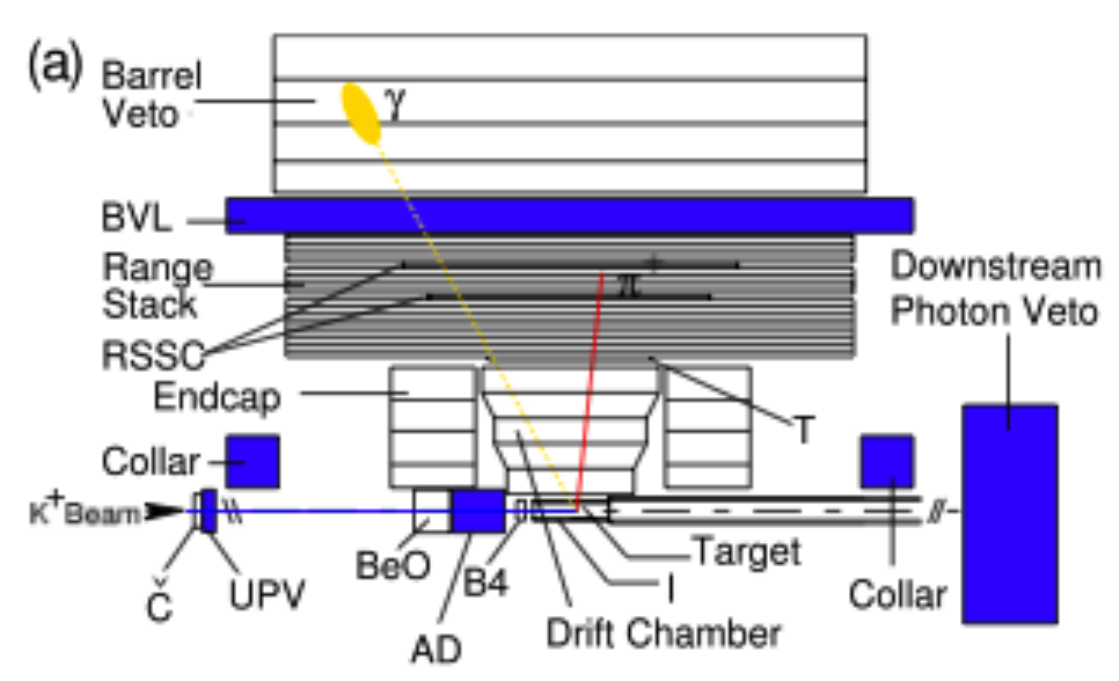}
\includegraphics[width=0.43\columnwidth]{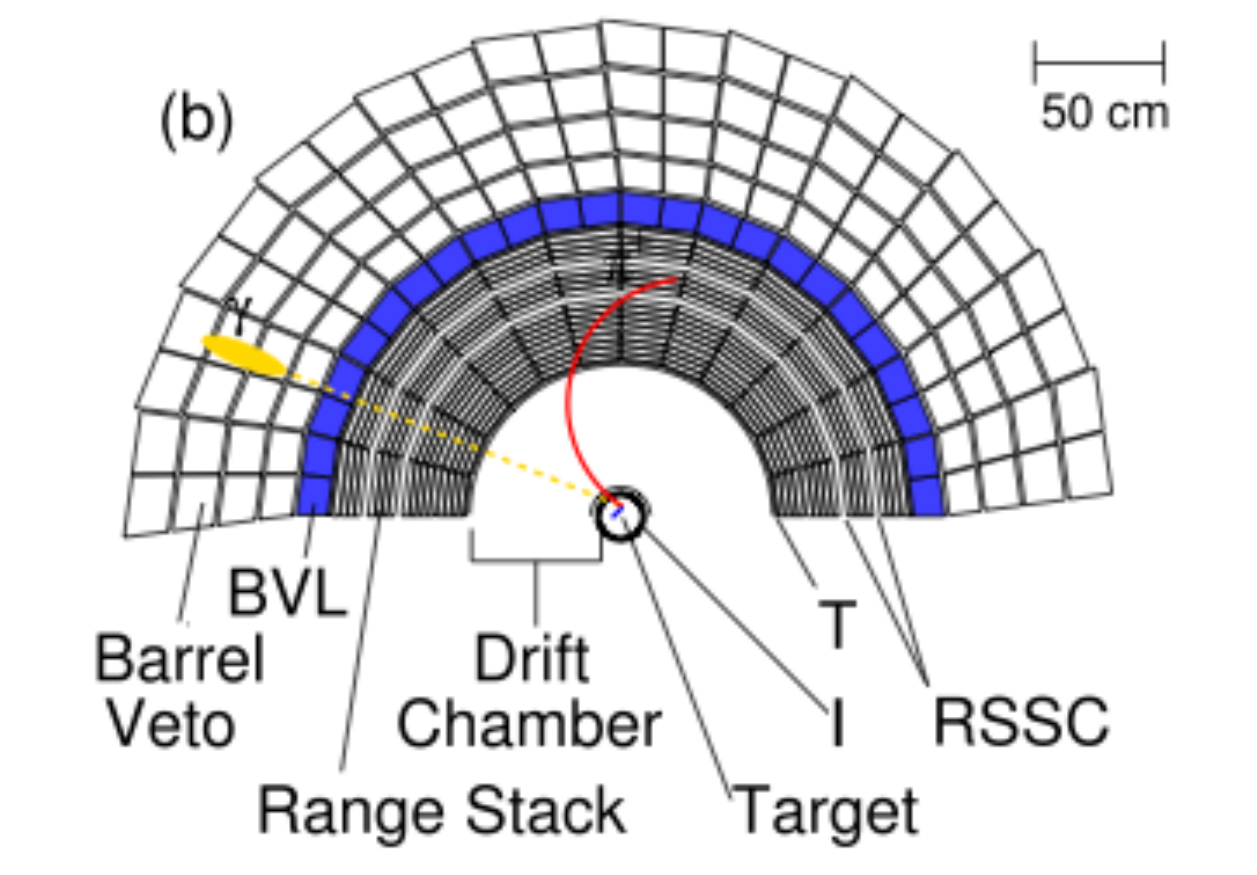}
\caption{Schematic side (a) and end (b) views of the upper half of the
BNL E949 detector. An incoming $K^+$ traverses all the beam instruments,
stops, and decays. The outgoing $\pi^+$ and one photon from 
$\pi^0 \rightarrow \gamma\gamma$ are also shown.}
\label{fig:e949det}
\end{figure}

The ORKA detector design will be based on that of BNL E949, a schematic 
of which is shown in Fig. 
\ref{fig:e949det}. Incoming $K^+$ particles are tracked, stop, and decay 
at rest in the scintillating-fiber
stopping target. The momentum of the $\pi^+$ 
from $\kpnn$ decay is 
analyzed in the drift chamber. The $\pi^+$ stops in the range stack 
where its
range and energy are measured and its decay products $\pme$  
are detected. A straw chamber in the range stack provides
additional information on the $\pi^+$ position. An extensive veto system 
provides 4$\pi$ photon veto coverage. The entire detector is immersed in
a 1 T solenoidal magnetic field along the beam direction.

\begin{figure}
\centering
\includegraphics[width=0.45\columnwidth]{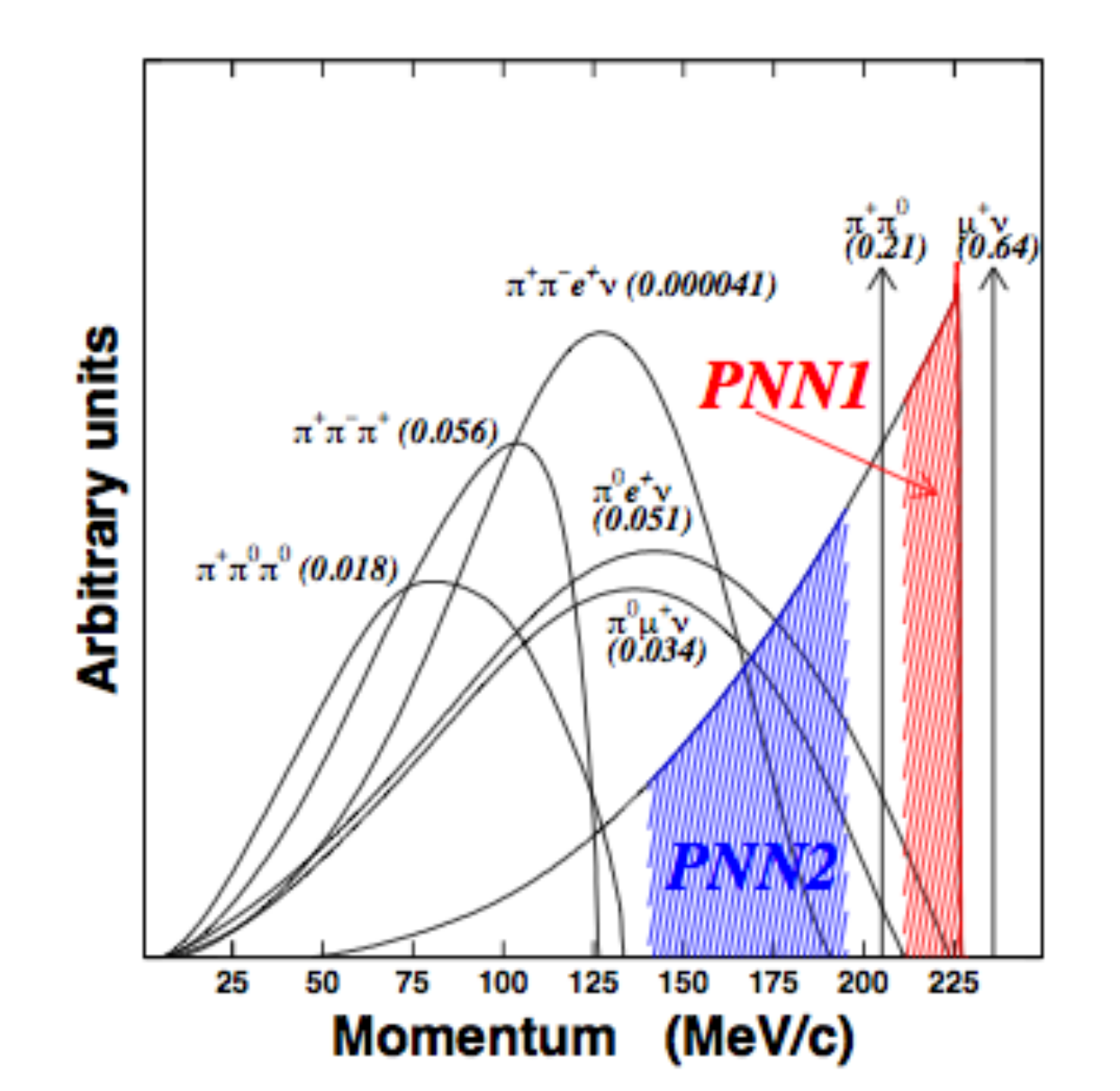}
\includegraphics[width=0.45\columnwidth]{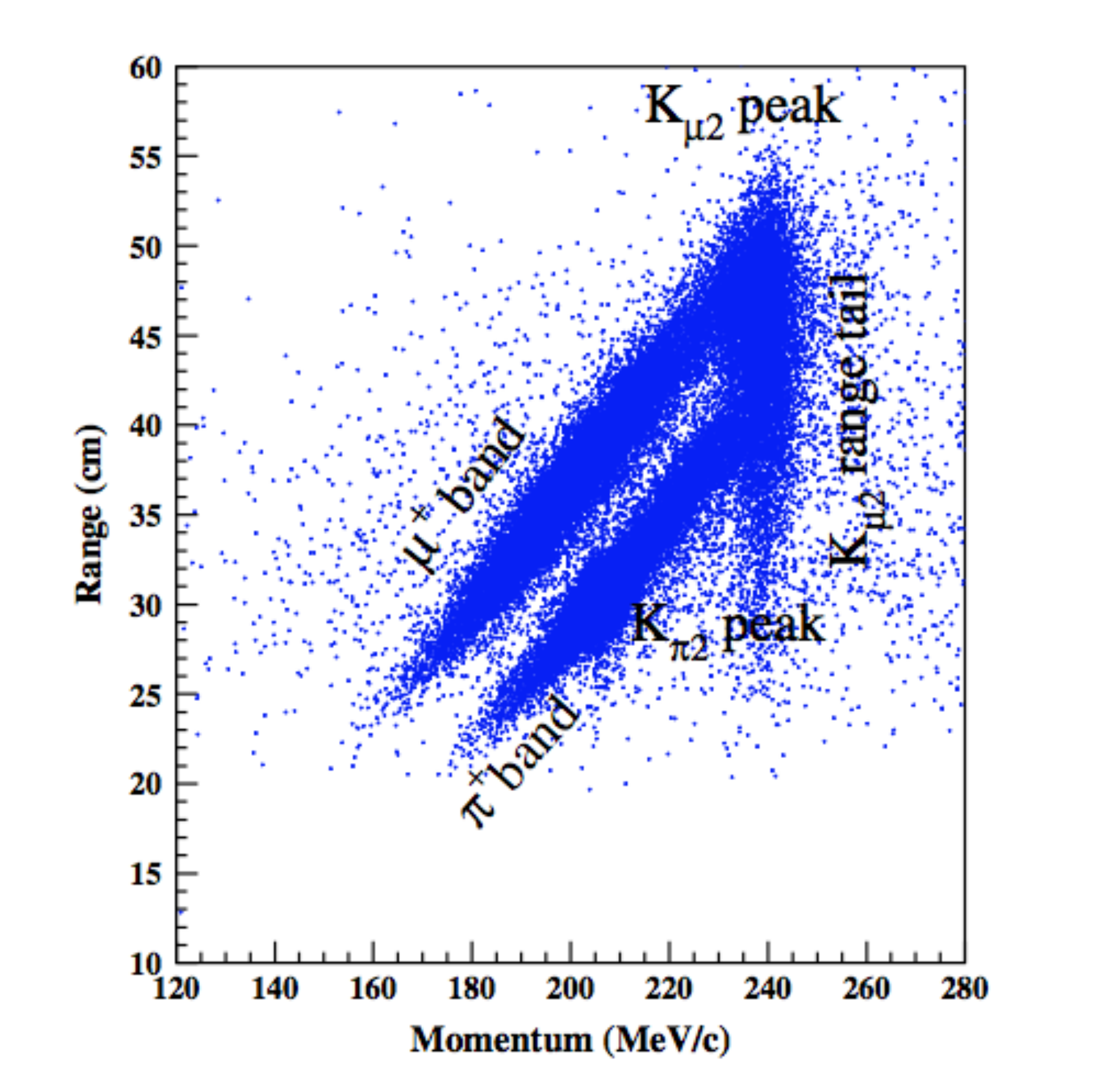}
\caption{(left) $\pi^+$ (or $\mu^+$) momentum distributions for several $K^+$ decay
modes. The respective branching ratios are shown in parentheses. The signal
regions used for the E787 and E949 $\kpnn$ branching ratio analyses (PNN1
and PNN2) are indicated.
(right) Range in plastic scintillator vs. momentum for charged particles
accepted by the $\kpnn$ PNN1 trigger. The concentrations of events
resulting from two-body decays are labeled ``$K_{\pi2}$ peak'' and
``$K_{\mu2}$ peak.'' The decays $K^+ \rightarrow \pi^0\mu^+\nu_{\mu}$ and
$K^+ \rightarrow \mu^+\nu_{\mu}\gamma$ contribute to the muon band.
The pion band results from $K^+ \rightarrow \pi^+\pi^0\gamma$ decay
in which the $\pi^+$ scatters in the target or range stack and beam
$\pi^+$ that scatter in the target.}
\label{fig:sigspec}
\end{figure}

Measurement of the $\kpnn$ branching ratio at the 10$^{-10}$ level 
is challenging because the signal, consisting only of the $\pme$
decay chain, is poorly defined and the background, 
primarily from
$K^+$ decays with branching ratios as much as ten orders of magnitude
larger than $\kpnn$ decay, is large and difficult to distinguish from the signal.
Figure \ref{fig:sigspec} (left panel) shows the momentum spectrum of the outgoing 
charged particle for
$\kpnn$ signal and important $K^+$-decay 
background sources. To suppress background
to a sufficiently low level, the detector must have excellent $\pi^+$
particle identification to reject background from $\kmtwo$ and 
$\kmtwo\gamma\gamma$ decays, highly efficient 4$\pi$ solid-angle 
photon veto coverage to
reject background from $\kptwo$ decays, and a $K^+$ identification system
to remove beam-related background.

Figure \ref{fig:sigspec} (left panel) 
also shows two kinematic regions that can be used to
search for $\kpnn$ decay in the data; these regions are chosen to avoid
background. In the PNN1 region, the $\pi^+$ momentum lies between 
the $\pi^+$ momentum from $\kptwo$ decay and the $\pi^+$ kinematic limit
of 227 MeV/$c$. The $\mu^+$ momentum from $\kmtwo$ decay lies above the
$\pi^+$ kinematic limit. 
In the PNN2 region, the $\pi^+$ momentum is below
that from $\kptwo$ decay, but analysis of this region is complicated by 
having additional sources of background from three-body
decays of the $K^+$. 
Beam related background,
in which a $K^+$ decays
in flight, a beam $\pi^+$ scatters in the stopping target,
or a $K^+$ undergoes charge exchange,
 must also be considered. Figure \ref{fig:sigspec} (right panel) shows the range and momentum of the charged 
particle for all events passing the PNN1
trigger. Background from $\kptwo$, $\kmtwo$, various 3-body decays,
and scattered pions are identified.

\begin{figure}
\centering
\includegraphics[width=0.9\columnwidth]{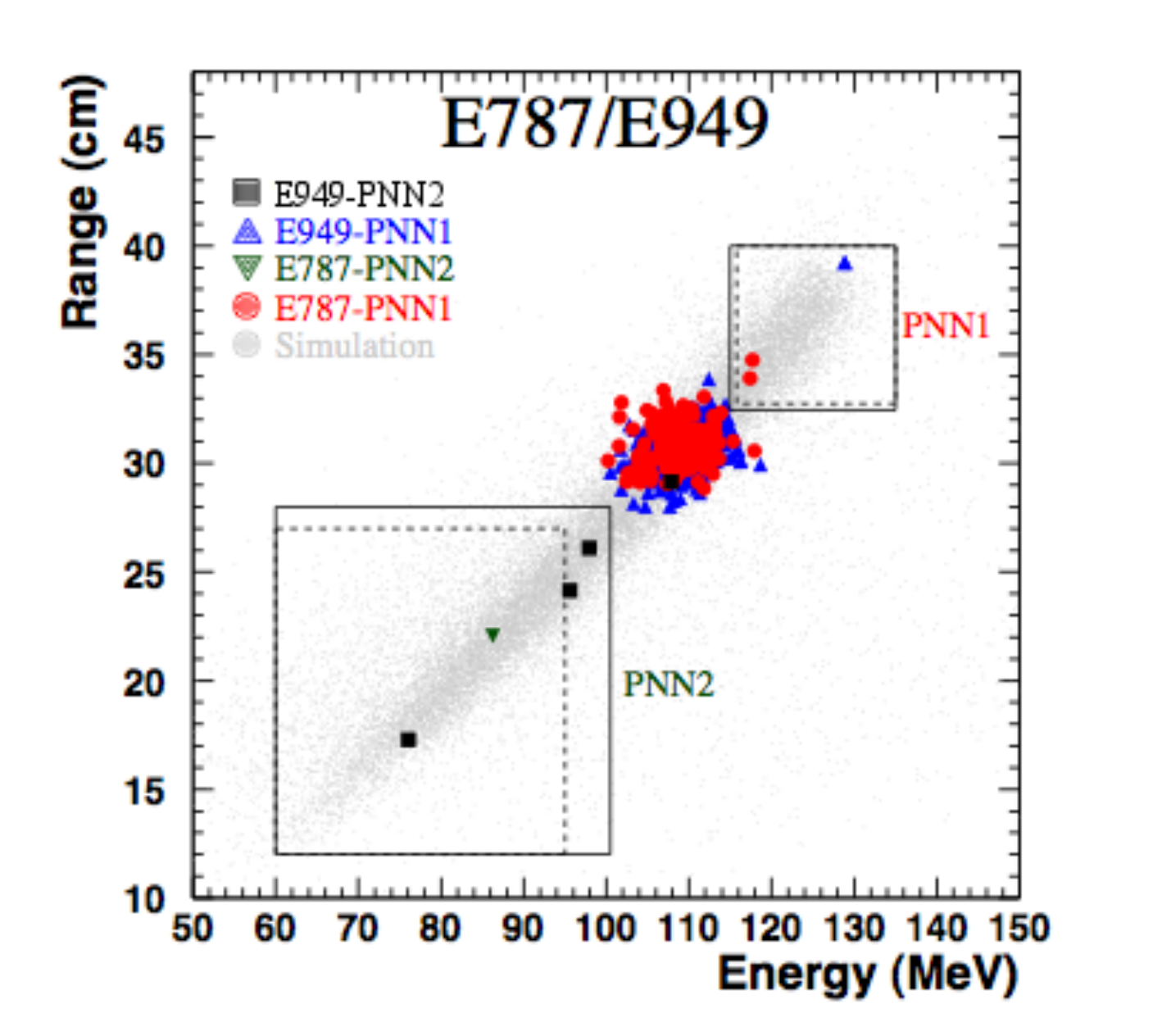}
\caption{Range vs. kinetic energy for events satisfying
all other selection criteria in BNL
E787/E949. The two signal regions, PNN1 and PNN2, are shown by
the dashed line (E787) and the solid line (E949). Events near E = 108 MeV
are background from $\kptwo$ decays which are not removed by photon veto
requirements. The light points represent the expected distribution
of $\kpnn$ events from simulation.}
\label{fig:signal}
\end{figure}

Data from BNL E787/E949 are shown in Fig. \ref{fig:signal}.
Data in the PNN1 and PNN2 regions were analyzed separately. The analysis
was done using a blind signal region. The final background estimates
were obtained from different data samples than were used to determine the 
selection criteria, using 2/3 and 1/3 of the data
respectively, and the background level was evaluated based on information from
outside the signal region.The expected number of
background events was $<<1$ in PNN1 and $\sim$1 in PNN2.
As seen in Fig. \ref{fig:signal}, a total of seven candidate
$\kpnn$ events were identified.

\subsection{ORKA Sensitivity}
\label{sect:orkaexpt}
The ORKA experiment will have a sensitivity $\sim$100 times
that of BNL E949. The beam line will provide
one factor of ten, while increases in detector acceptance provide
the other order of magnitude.

The ORKA experiment intends to extract a 95-GeV
beam from the MI ring onto a production target, select a 600-MeV/$c$ $K^+$
beam from that target, and bring the kaons to rest at the center of the
detector. The FNAL MI provides $\sim$25\% fewer protons per spill than the
BNL beam. However, ORKA's kaon beam line provides a factor of $>$4 times the
number of
kaons per proton from a longer target, increased angular acceptance, and
increased momentum acceptance. The beam-line design also allows a 40\%
increase in the fraction of  kaons that survive until the stopping target
and an additional factor of 2.6 in the fraction of
kaons that come to rest in the stopping target. The detector will operate
at a higher instantaneous rate, which leads to $\sim$10\% more losses from
vetoed events. Overall, the ORKA beam line will provide ten times more 
$\kpnn$ events than BNL E949. 

The ORKA detector, the design of which is modeled closely on that of
BNL E949 (see Fig. \ref{fig:e949det}),
will have a factor of ten greater acceptance than the BNL
E949 detector. Most of the increases in acceptance are incremental;
a variety of improvements to the detector design each provide 
acceptance increases of 6-60\%. The
largest change in acceptance is a factor of 2.2 increase in the 
$\pme$ acceptance, which
is the result of several features of the ORKA detector.
The ORKA range stack has increased segmentation relative to the range stack
in BNL E949; this will reduce
losses from accidental activity and improve $\pi/\mu$ particle identification.
ORKA will use higher quantum efficiency photo-detectors and/or better optical
coupling to increase the scintillator light yield, which will improve $\mu$
identification. ORKA will also employ a modern, deadtime-less DAQ and trigger
so that losses from online $\pi/\mu$ particle identification will be
eliminated. The irreducible losses to $\kpnn$ acceptance are from necessary
cuts on the measured $\pi^+$ and $\mu^+$ lifetimes and from $\mu^+$ and
$e^+$ particles that do not deposit enough energy to be detected. In the 
PNN1 region, the $\pme$ acceptance in ORKA is
estimated to be $\sim$78\%.

\begin{figure}
\centering
\includegraphics[width=0.9\columnwidth]{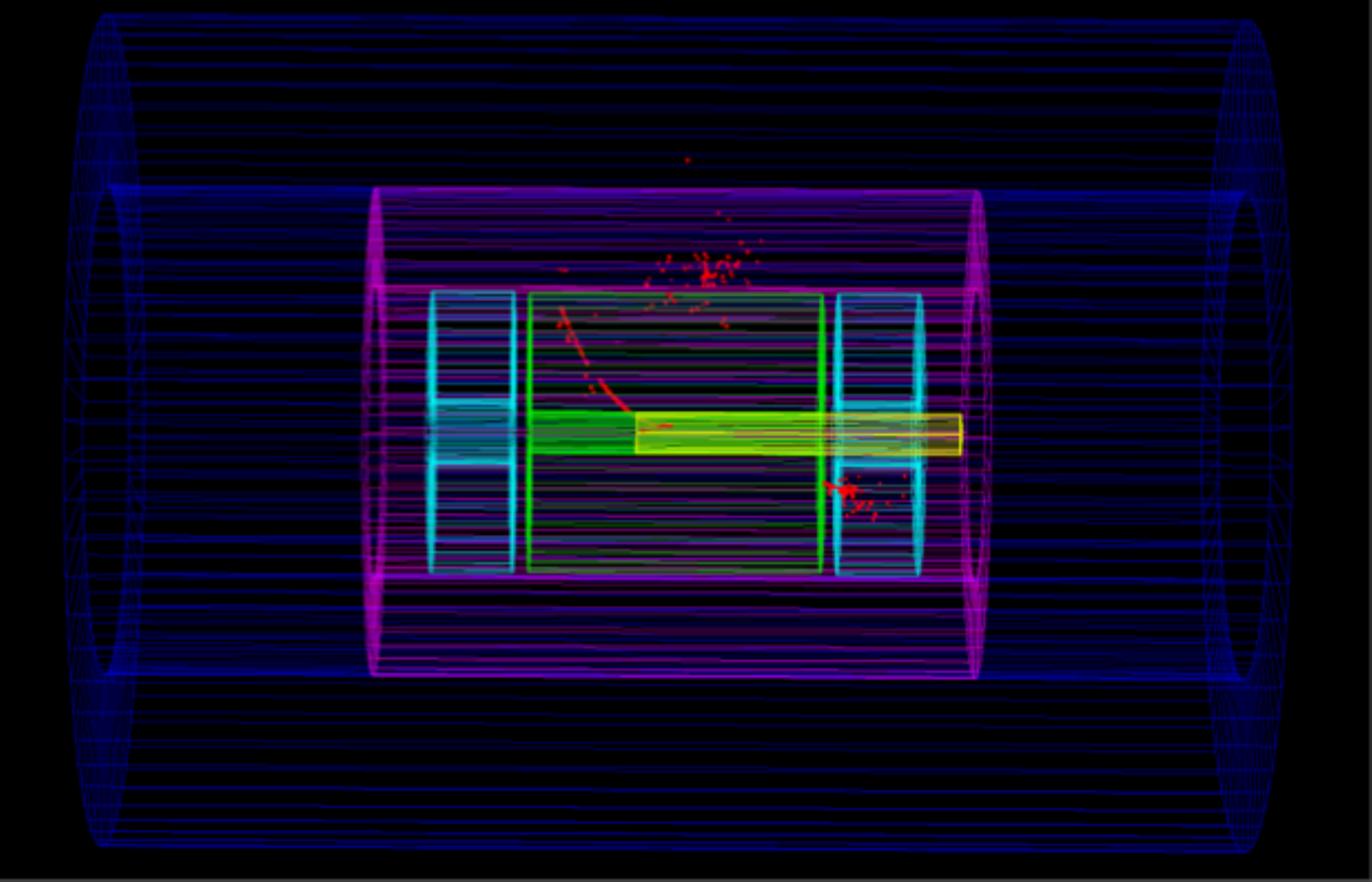}
\caption{Event display from ILCroot simulation of the ORKA detector. The target
is shown in yellow, the drift chamber in green, the range stack in pink, the barrel
veto in blue, and the veto end-caps in cyan. The simulated event is a $\kptwo$ decay.
The two photon clusters from $\pi^0$ decay and the $\pi^+\rightarrow\mu^+$ tracks
are visible.}
\label{fig:orkadetsim}
\end{figure}

Preliminary simulation of the ORKA detector for optimization of the detector
design is underway. Figure \ref{fig:orkadetsim} shows an ILCroot\cite{ilcroot}
event display of the ORKA detector in which a $\kptwo$ decay has been simulated.

\begin{figure}
\centering
\includegraphics[width=0.9\columnwidth]{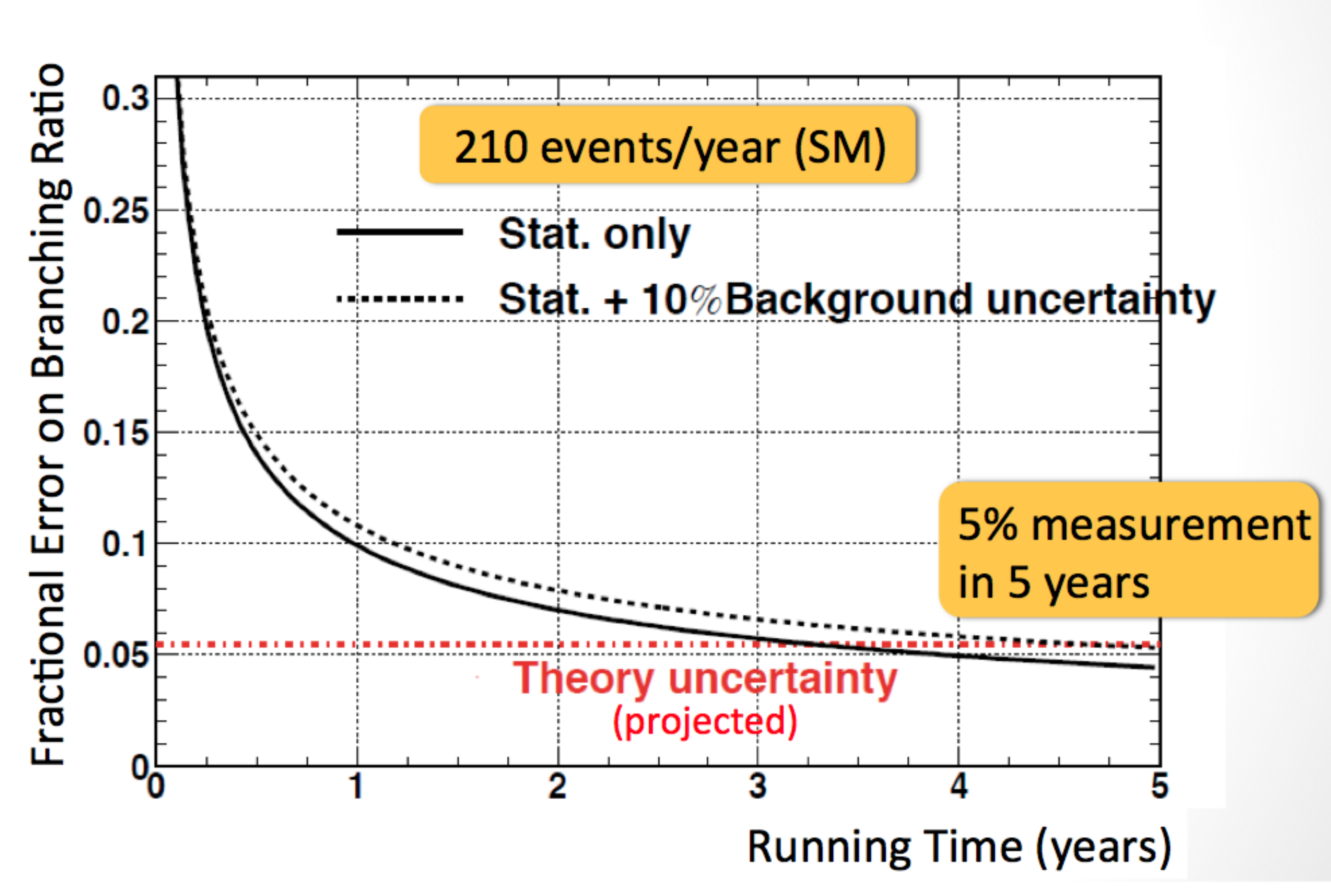}
\caption{The fractional uncertainty in the ORKA measurement
of the $\kpnn$ branching ratio as a function of running time. The
solid (dashed) curve is the sensitivity assuming no (10\%) background
uncertainty. The flat dashed line is the uncertainty in the theoretical
prediction of the $\kpnn$ branching ratio, excluding uncertainties in the
CKM matrix elements.}
\label{fig:sensitivity}
\end{figure}

The ORKA experiment will detect $\sim$210 $\kpnn$ events per year at the
Standard Model level. Figure \ref{fig:sensitivity} shows the fractional
error on the $\kpnn$ branching ratio as a function of running time. ORKA
expects to reach the projected theoretical uncertainty on the 
$\kpnn$ branching
ratio after taking data for about five years.

\subsection{Physics Breadth in ORKA}
\label{sect:otherphys}
The ORKA detector is highly optimized for detection of $\kpnn$ decay; the
signal for this decay is basically $K^+ \rightarrow \pi^+$ plus missing 
energy. In the case of $\kpnn$ decay, the missing energy is neutrinos,
but other forms of missing energy can be sought by observing the shape
of the $\pi^+$ spectrum. For this reason, ORKA will have sensitivity 
to many modes of considerable interest. It is also true that a very large
number of stopped kaons will decay in the ORKA detector. The sheer volume
of data may facilitate measurements even in decay modes for which the detector
is not optimized.

One possible source of missing energy is the case in which a single unseen
particle recoils from the $\pi^+$. In this case, the $\pi^+$ spectrum is
a peak and its width is determined by the detector resolution. Theoretical
models exist in which the unseen $X^0$ is
a familon\cite{familon}, an axion\cite{axion}, a light scalar
pseudo-Nambu-Goldstone boson\cite{ngboson}, a sgoldstino\cite{sgoldstino},
a gauge boson ($A^{\prime}$) corresponding to a new U(1) gauge symmetry 
\cite{u11,u12}, or
a light dark-matter candidate\cite{dm1,dm2,dm3}. 


ORKA will have sensitivity to many other interesting decay modes. These
include: $K^+ \rightarrow \pi^+\pi^0\nu\bar{\nu}$, 
$K^+\rightarrow\pi^+\gamma$, $K^+\rightarrow\mu^+\nu_{heavy}$, 
$K^+\rightarrow\pi^+\gamma\gamma$, $\pi^0\rightarrow\nu\bar{\nu}$,
and $\pi^0\rightarrow\gamma X^0$. There are other possible searches, such as
$K^+\rightarrow \pi^+ A^{\prime}, A^{\prime} \rightarrow e^+e^-$,
for which the ORKA detector is not optimized, but might still have 
sensitivity because of the large numbers of stopped-kaon
decays.

ORKA will also be able to make precise measurements of important kaon
branching ratios and fundamental parameters. $R_K = \Gamma(K^+ \rightarrow
e^+\nu_e)/\Gamma(K^+ \rightarrow \mu^+\nu_{\mu})$ has been measured to 0.4\%
by NA-62\cite{rkpub_na62}. ORKA will be able to measure this ratio 
with <0.1\% statistical precision; study of the expected systematic 
uncertainty is still needed. Other fundamental kaon measurements accessible 
to ORKA include
the $K^+$ lifetime and the 
$B(K^+\rightarrow\pi^+\pi^0)/B(K^+\rightarrow\mu^+\nu_{\mu})$ ratio.

\subsection{Status of ORKA}
\label{sect:orkastatus}
The ORKA collaboration submitted a proposal\cite{orkaproposal} to 
FNAL in November 2011.
The experiment received Stage 1 approval
from the FNAL directorate in December 2011. The planned site for the 
experiment is B0, so that ORKA can re-use the CDF solenoid, cryogenics,
and infrastructure. The ORKA detector is being designed to fit inside the
CDF solenoid. Decommissioning of
CDF in a way that facilitates eventual use of CDF hall by ORKA is underway; 
this effort
is being funded in the FY13 FNAL budget. Plans are being made and beam-line
elements are being identified for a beam line to
transport protons from the Main Injector extraction point to B0 in the 
Tevatron tunnel. A preliminary design for the kaon beam line from the
production target to the detector has been simulated and is being refined.
Research and development on detector design has begun. 
The ORKA collaboration consists of 17 institutions from six
countries.
Because ORKA is a
fourth generation detector, building on experience from BNL E787/E949, an
aggressive construction schedule is thought possible; the
collaboration expects to begin taking data within five years of
funding approval. The ORKA collaboration is working with the DOE to determine
when funding will be available.

\section{Summary}
\label{sect:summary}
Precision measurement of the $\kpnn$ branching ratio is an opportunity
to search for the effects of new physics at and beyond the mass scale
accessible by direct searches. The ORKA experiment will use proven
accelerator technology,
detector technology, and experimental techniques to measure the $\kpnn$
branching ratio with an uncertainty of $\sim$5\%, allowing 
$5\sigma$-level detection 
of deviations from the Standard Model as small as 35\%. 
The ORKA experiment will
make use of protons from the FNAL Main Injector and will be sited
at B0 at Fermilab. The collaboration is currently 
conducting detector R\&D in preparation for ORKA.

\bibliographystyle{ieeetr}
\bibliography{etw_orka_bib}

%

\end{document}